 \documentstyle[psfig,twocolumn,prl,aps,amssymb]{revtex}
 \begin{document}
 \wideabs{
 \draft

 \date{\today} \title{Possible Pairing-Induced Even-Denominator Fractional
 Quantum Hall Effect
 in the Lowest Landau Level}
 \author{V.W. Scarola$^1$, J.K. Jain$^1$, and E.H. Rezayi$^2$}
 \address{$^1$Department of Physics, 104 Davey Laboratory, The Pennsylvania State
 University, University Park, Pennsylvania 16802}
 \address{$^2$Department of Physics, California State University, Los Angeles,
 California 90032}
 \maketitle

 \begin{abstract}
 We report on our theoretical investigations that
 point to the possibility of a fractional quantum Hall effect with partial spin
 polarization
 at $\nu=3/8$.  The physics of the incompressible state proposed here
 involves p-wave pairing of composite fermions in the spin reversed sector.
 The temperature and magnetic field regimes for the realization of this state are
 estimated.
 \end{abstract}
 }

 The absence of fractional quantum Hall effect at the simplest
 even denominator fraction, $\nu=1/2$, continued to be an enigma for a decade
 after the
 discovery of the fractional quantum Hall effect (FQHE),\cite{Tsui}
 but found a natural explanation \cite{Halperin,Zhang}
 in the framework of the  composite fermion (CF) theory of the
 FQHE.\cite{Jain1,CFbook}
 The sequence $\nu=\frac{n}{2n\pm 1}$, corresponding to the integral quantum Hall
 effect of composite fermions carrying two vortices (denoted by $^2$CFs) at
 $\nu^*=n$,
 converges to $\nu=1/2$ in the limit $n\rightarrow \infty$.
 At least for a model of non-interacting composite
 fermions, a gapless Fermi sea of composite fermions is obtained here, for
 which good experimental support exists.\cite{half} A FQHE at $\nu=1/2$ is not
 ruled
 out in principle, though; it may occur if, due to the residual inter-CF
 interaction,
 the CF Fermi sea should become unstable into an incompressible state as the
 temperature is lowered.  There is, however, no experimental evidence at present
 for such
 an instability at $\nu=1/2$.

 One might expect similar physics at the
 half filled {\em second} Landau level (LL), $\nu=5/2$, but a
 FQHE is observed here instead.\cite{Willett,Pan}
 The most promising proposal for the physical origin of the FQHE at $\nu=5/2$ is
 based on a p-wave pairing of composite
 fermions,\cite{Moore,Greiter,Morf,Park,Haldane3,Haldane2,Scarola}
 described by a BCS-like Pfaffian wave function of Moore and
 Read\cite{Moore}.  In spite of the intuitive appeal and theoretical support of
 this idea,
 further experimental tests of its consequences are crucial for its
 establishment.\cite{Willett2}  The difference between $\nu=1/2$ and $\nu=5/2$
 lies
 in the interaction matrix elements, i.e. the Haldane
 pseudopotentials.\cite{Haldane}
 A strong short range repulsion between electrons produces a Fermi sea of
 composite
 fermions, but when the interaction becomes weakly repulsive at short distances,
 as in the second LL, it produces an effective
 {\em attractive} interaction between composite fermions, causing a pairing
 instability of the
 CF Fermi sea \cite{Scarola}. For an attractive interaction between electrons,
 either
 a charge-density-wave (CDW) phase \cite {Koulakov} or a spin-singlet FQHE
 state\cite{HR}
 becomes relevant, depending on parameters.  The p-wave pairing between composite
 fermions
 thus seems to be favored when the interparticle interaction is {\em weakly}
 repulsive
 at short distances.

\begin{figure}
 \centerline{\psfig{figure=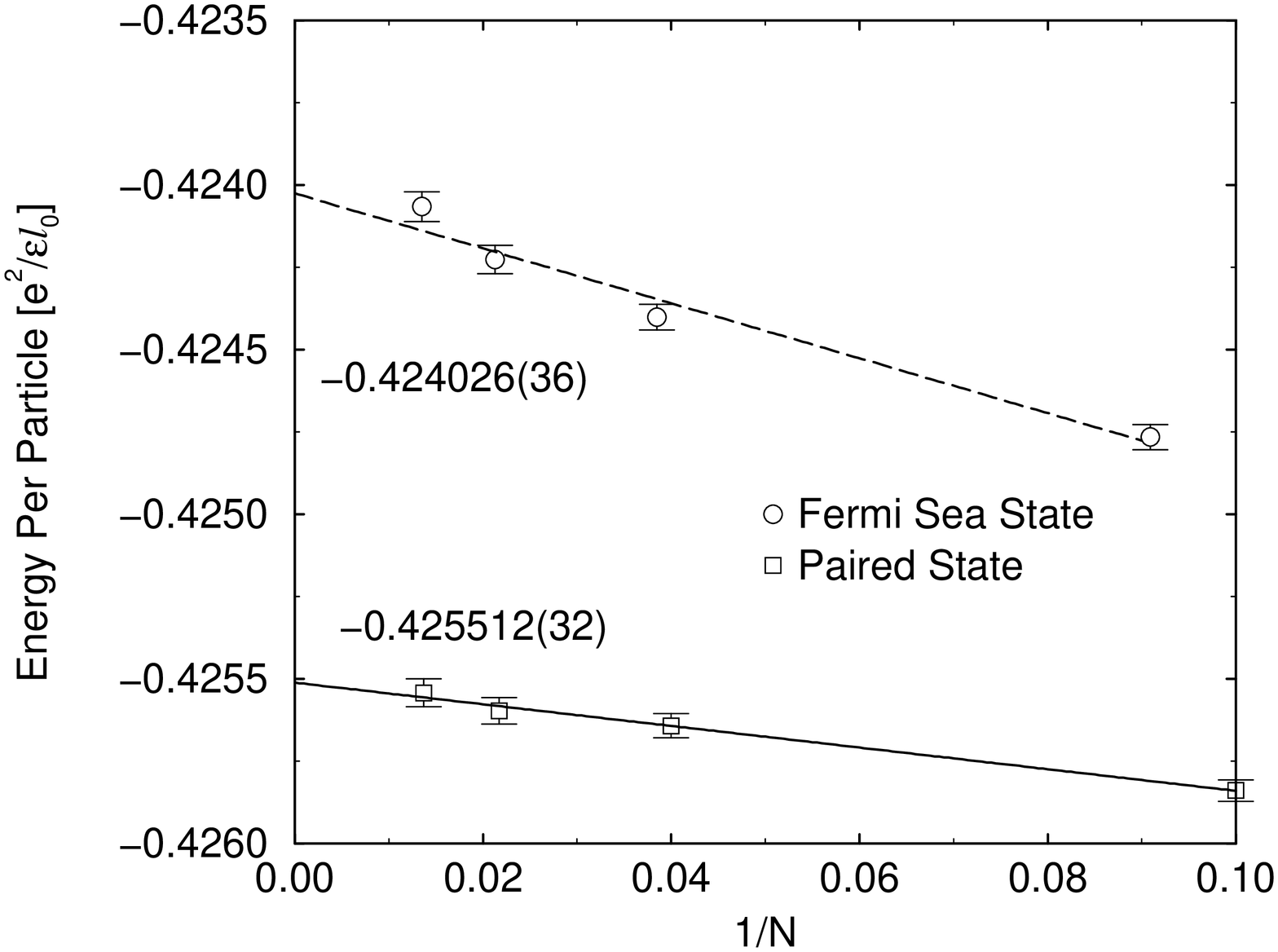,width=3.5in,angle=-90}
 \put(-120,90){\psfig{figure=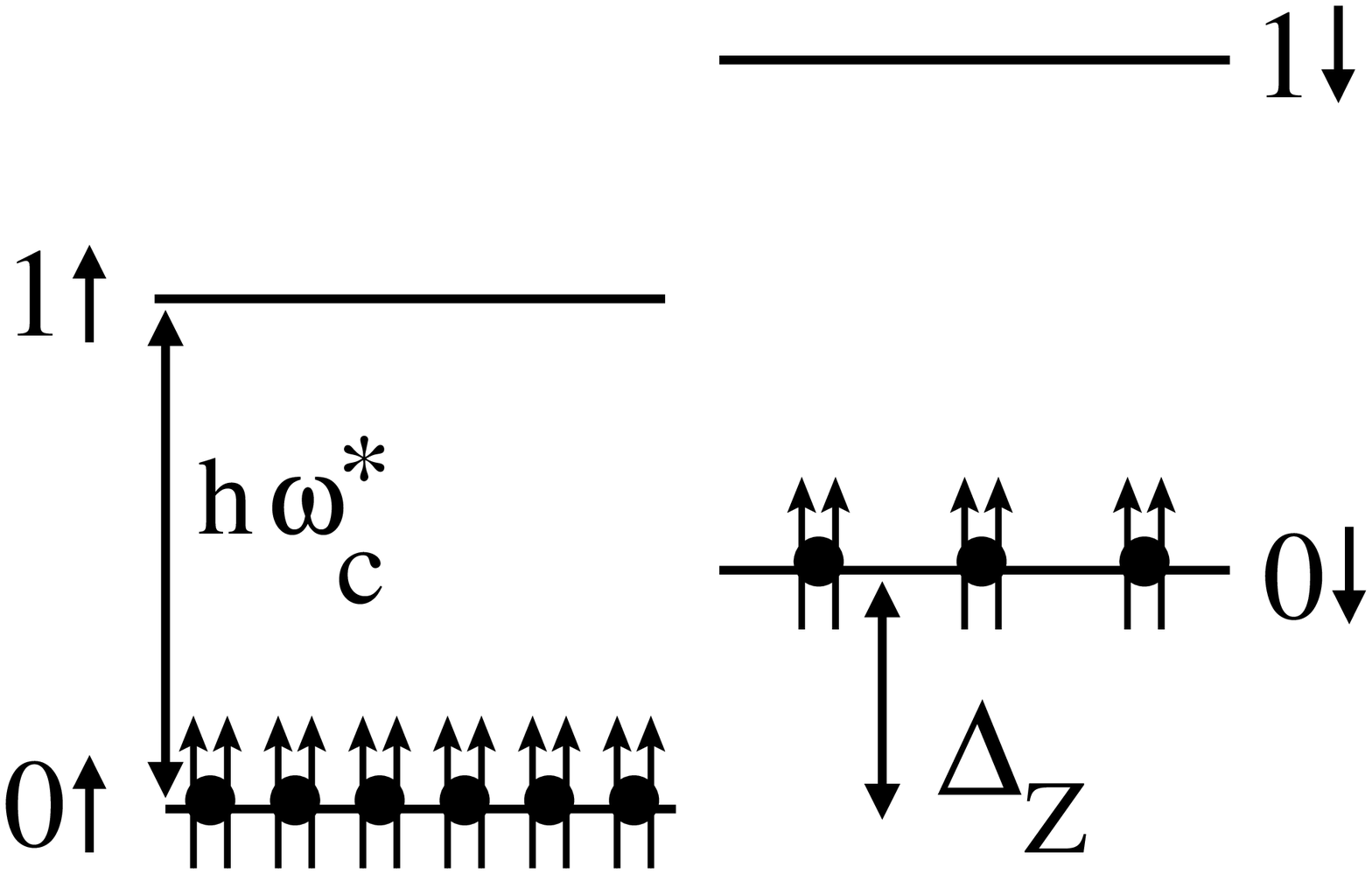,width=1.25in,height=1.25in,angle=0}}}
 \caption{The composite fermion system is considered at $\nu^*=1+1/2$ with
 the 0$\uparrow$ $^2$CF LL is fully occupied and the 0$\downarrow$ $^2$CF LL half
 filled, as shown schematically in the inset. The CF LL spacing is
 $\hbar\omega_{c}^{*}$ and the
 Zeeman splitting is denoted by $\Delta_{Z}$.
 This system corresponds to a partially polarized state at $\nu=3/8$.
 Two states are considered in which the
 composite fermion in the 0$\downarrow$ $^2$CF LL either form a $^4$CF Fermi sea
 or
 a $^4$CF paired state. The energies of these states are shown as a function of
 $1/N$,
 $N$ being the total number of
 particles, in units of $e^2/\epsilon l_0$, where $l_0=\sqrt{\hbar
 c/eB}$ is the magnetic length and $\epsilon$ is
 the dielectric constant of the background material.
 The error bars reflect the statistical uncertainty in our Monte
 Carlo calculation.
 }
 \label{therm}
 \end{figure}

 An example of weakly interacting fermions
 is composite fermions themselves.  This raises the natural question if they
 could ever do what electrons do at $\nu=5/2$, namely put on (additional)
 vortices
 and pair up to produce FQHE.  After a careful consideration of a wide range of
 possibilities, we
 have concluded that the best candidate is at CF filling of $\nu^*=1+1/2$, when
 the
 0$\uparrow$ Landau level of $^2$CFs is fully occupied
 and the 0$\downarrow$ $^2$CF Landau level is half filled, as shown
 schematically in the inset of Fig.~(\ref{therm}).  This corresponds to a
 partially polarized state at $\nu=3/8$.  The reason why this system is a good
 candidate
 for pairing is because the CF-CF interaction here is weakly repulsive at short
 distances,
 the origin of which can be understood
 intuitively following an argument by Nakajima and Aoki.\cite{Aoki}  The Haldane
 pseudopotentials for composite fermions, $V_m^{CF}$, are expected to be
 related to the electron pseudopotentials in the lowest LL, $V^{elec}_m$,
 approximately
 according to $V_m^{CF}\propto V^{elec}_{m+2p}$, because a capture of $2p$
 vortices
 by electrons shifts the relative angular momentum of any pair by $2p$ units.
 The strong short-range repulsion is thus eliminated when electrons transform
 into
 composite fermions.\cite{Comment2} Below we describe our investigations that
 indeed
 support the possibility of a p-wave pairing of composite fermions in the spin
 reversed sector.

 The spatial part of the wave function of the electronic state at $\nu=3/8$ is
 written
 as:
 \begin{equation}
 \Psi_{\frac{3}{8}}=J^2\phi_{1}^{\uparrow}[\{w_r\}]\phi_{1/2}^{\downarrow}[\{z_i\}]
 \label{wf}
 \end{equation}
 \begin{eqnarray}
 J^2=\prod_{r<s}^{N_\uparrow}(w_r-w_s)^2\prod_{i<j}^{N_\downarrow}(z_i-z_j)^2
 \prod_{i,r}^{N_\downarrow,N_\uparrow}(w_r-z_i)^2
 \end{eqnarray}
 where $w_r=x_r-iy_r$ and $z_j=x_j-iy_j$ refer to the coordinates of the
 electrons with
 up and down spins, respectively.
 The full wave function is written by multiplying by the appropriate spin part
 followed by
 antisymmetrization.  It has spin polarization
 $(N_\uparrow-N_\downarrow)/(N_\uparrow+N_\downarrow)=1/3$, and can be shown to
 be an
 eigenstate of the total spin\cite{MacDonald,Park} with $S=S_z$.

 The factor $\phi_1^{\uparrow}[\{w_r\}]$ is the wave function for
 the completely occupied lowest Landau level.
 Different states at $\nu=3/8$ correspond to different choices for
 $\phi_{1/2}^{\downarrow}[\{z_i\}]$
 For the Fermi sea state at $\nu=3/8$, $\Psi_{\frac{3}{8}}^{FS}$, we take in
 Eq.~(\ref{wf})
 \begin{equation}
 \phi_{1/2}^{\downarrow}[\{z_i\}]={\cal P}_{LLL}
 \prod\limits_{j<k}(z_j-z_k)^2\phi_{\infty}^{\downarrow}[\{z_i\}]
 \end{equation}
 where $\phi_{\infty}^{\downarrow}$ is the Fermi sea wave function of electrons
 in
 zero magnetic field and ${\cal P}_{LLL}$ is the lowest Landau level projection
 operator.
 The Pfaffian state, $\Psi_{\frac{3}{8}}^{Pf}$, is obtained with the
 choice \cite{Moore}
 \begin{equation}
 \phi_{1/2}^{\downarrow}[\{z_i\}]=\prod\limits_{j<k}(z_j-z_k)^2 Pf(M^\downarrow)
 \end{equation}
 where $Pf(M^{\downarrow})$ is the Pfaffian of the
 $N_{\downarrow}\times N_{\downarrow}$ antisymmetric matrix $M^{\downarrow}$
 with components $M^{\downarrow}_{j,k}=(z_j-z_k)^{-1}$, defined as
 $Pf(M^{\downarrow})\propto A[M_{12}M_{34}...M_{N_\downarrow-1,N_\downarrow}]$,
 $A$ being the antisymmetrization operator.  $Pf(M^\downarrow)$ is a
 real space BCS wave function and $\phi_{1/2}^{\downarrow}$ can therefore be
 viewed as a p-wave paired quantum Hall state of composite fermions.

 These states not only have mixed spin, but also an admixture of different
 flavors of composite
 fermions,\cite{Park2} those carrying 2 and 4 vortices, called $^2$CFs and
 $^4$CFs, respectively.
 In the first case, $^2$CFs capture two additional vortices to convert into
 $^4$CFs, which
 effectively
 experience no magnetic field and form a Fermi sea.  In the second case, the
 $^2$CFs capture two
 additional vortices to convert into $^4$CFs,
 which form pairs; a gap opens up due to pairing and FQHE results.
 The wave functions above can be interpreted as describing
 Fermi sea and paired states of spin-down $^4$CFs in the background of spin-up
 $^2$CFs.

 To check which state is energetically superior, we use the explicit analytical
 expressions
 for the lowest LL projected wave functions for composite fermions
 \cite{JK,CFbook} and perform $2N$ dimensional integrals using Monte Carlo
 to obtain their Coulomb energy.  We map the states onto the surface of a sphere
 \cite{Haldane} and calculate the energy of each state in the
 thermodynamic limit using a least squares fit.  Fig.~(\ref{therm})
 shows that while both the paired state and the Fermi sea have energies quite
 comparable to
 those of the
 fully polarized states at 1/3 and 2/5 (-0.4098 and -0.4328 $e^2/\epsilon l$,
 respectively \cite{JK,CFbook}), the former is favored over the latter.
 The energy difference is reduced by approximately an order of magnitude as
 compared to the
 analogous difference at 5/2,\cite{Park} indicative of a substantially weaker
 interaction between
 the composite fermions.

 Though consistent with pairing within the spin-down sector, the variational
 nature of the preceding result does not rule out other ground states.  One way
 to ascertain
 the validity of $\Psi_{\frac{3}{8}}^{Pf}$ would be to compare it with exact
 results
 for small systems.  Unfortunately,
 the smallest system size for this state is 10 particles in the spherical
 geometry, where the Hilbert
 space is already prohibitively large for an exact diagonalization study.
 To make progress, we treat the filled 0$\uparrow$ $^2$CF Landau level as inert,
 and
 map the problem of composite fermions at $\nu^{*}=1+1/2$ into fermions at half
 filling.
 The advantage of this method is that only the reversed spin
 composite fermions in the 0$\downarrow$ $^2$CF-LL are considered explicitly,
 which reduces the size of the Hilbert space considerably.  Given the
 strongly correlated nature of the problem, the effective $^2$CF-$^2$CF
 interaction is complicated and is expected to
 contain two, three, and higher body terms.  In order to proceed,
 we neglect all but the two body term; i.e., we
 assume that the three and higher body terms do not cause any phase transition.
 This approximate treatment of the effective interaction between these composite
 fermions
 is the most serious limitation of our model.
 The physics of the problem suggests the following CF-CF interaction.
 Consider two electrons in the lowest LL.  Attaching  $2p$ vortices to each
 electron modifies
 the interaction in three ways:
 i)  The charge is reduced by a factor $(2p+1)^{-1}$.
 ii)  The effective magnetic field seen
 by CFs increases the magnetic length by a factor $(2p+1)^{\frac{1}{2}}$.
 iii) The model pseudopotentials have their relative angular momentum
 shifted by $2p$.
 This motivates the following model, similar to the one used previously
 for an investigation of the spin wave dispersion at $\nu=1/3$:\cite{Aoki}
 \begin{eqnarray}
 \frac{V^{elec}_{m^*}}{e^{*2}/\epsilon l^*}=\frac{V^{CF}_{m}}{e^2/\epsilon l_0}
 \end{eqnarray}
 where $m^*=m+2p$, $e^*=\frac{e}{2p+1}$, and $l^*=l_0\sqrt{2p+1}$.
 A comparison with the pseudopotentials obtained directly from the microscopic
 wave functions,
 following the method of Refs.~\cite{Quinn,Seung}, demonstrates that the above
 model is
 surprisingly accurate.  It should be noted that this model is valid only for
 spin-reversed
 composite fermions on top of the 1/3 state; in general, the more complicated
 method of
 Refs.~\cite{Quinn,Seung} must be used.\cite{Comment2}

 With the above effective interaction between the spin-down composite fermions,
 we carry out exact diagonalization to look
 for pairing correlations within the 0$\downarrow$ $^2$CF LL.
 To begin with, we find a uniform ($L=0$) ground state at the flux
 $2Q=2N_{\downarrow}-3$,
 which corresponds to the Pfaffian wave function,
 for $N_{\downarrow}=8$, 10, 14 and 16.\cite{comment}
 Next we compare the ground state to the Pfaffian wave function,
 which is obtained by diagonalizing an interaction (containing three body
 terms) for which the Pfaffian wave function is the only zero energy eigenstate.
 The overlaps between the ground state of the $V^{CF}$ interaction
 (which we identify with the 3/8 state here) and the Pfaffian wave function are
 given in Table I; for
 comparison, the overlaps between the ground state at 1/2 filling in the second
 electronic Landau level
 (identified with 5/2) and the Pfaffian are also given.  While the overlaps are
 not as
 decisively large as those between the filled CF-LL wave functions and the exact
 ground states at the principal filling
 factors,\cite{CFbook} they are significant, and overall support
 the interpretation of the 3/8 state as a paired state of composite fermions.

\begin{figure}
 \centerline{\psfig{figure=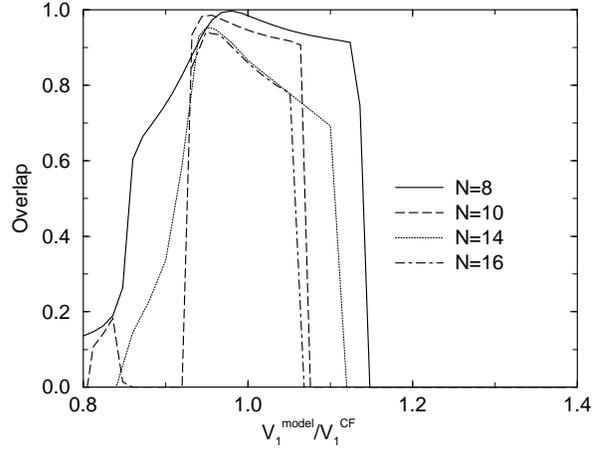,height=3.0in,angle=-90}}
 \caption{The overlap between the
 paired CF wave function of Moore and Read and the exact ground state of the
 model in which $V_1$ of the effective CF interaction is varied for $N=8$, 10, 14
 and 16
 particles.
 }
 \label{overlap}
 \end{figure}

 To address the issue of the robustness of the 3/8 paired state, we have
 investigated its
 evolution in a model in which the first odd pseudopotential, $V_1^{CF}$, is
 replaced by
 $V_1^{model}$.  As shown in Fig.~(\ref{overlap}), the
 state survives an $\sim$8\% change in $V_1$ in either direction.
 Based on the previous discussion, one would
 expect that on the large $V_1^{model}$ side of
 the ``paired" region ($V_1^{model}/V_1^{CF}\approx 1$),
 the Fermi sea has the lowest energy, whereas the stripe phase is likely
 on the small $V_1$ side.  We have confirmed this by comparing the energies of
 the
 three states following the method of Ref.~\cite{Seung}.
 This points to the interesting possibility that
 a transition from the paired state to stripes or Fermi sea may be driven by
 a change of parameters, as was suggested at 5/2 as well.\cite{Haldane2}
 It is noted that for the {\em fully} spin polarized state at $\nu=3/8$,
 where one must consider half filled 1$\uparrow$ $^2$CF Landau level, a similar
 model for the CF-CF interaction appears to indicate the stripe
 phase.\cite{Seung}

\begin{figure}
 \centerline{\psfig{figure=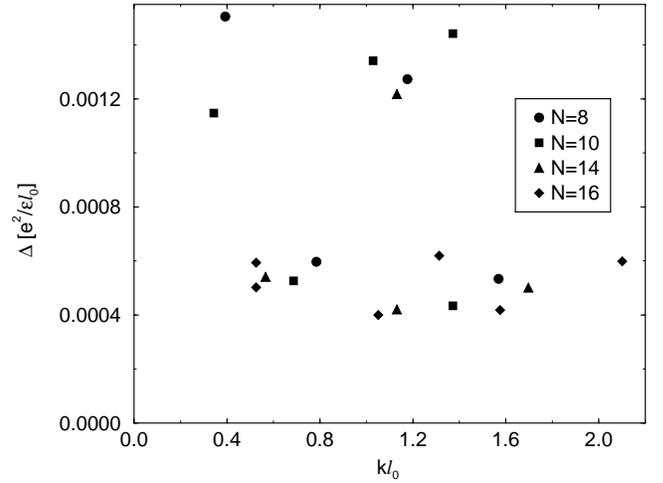,height=3.0in,angle=-90}}
 \caption{The low-energy spectrum
 for the CF model interaction.  The lowest energy states
 form a band separated from the continuum.  The dashed line is a guide
 to the eye.
 }
 \label{dispersion}
 \end{figure}

 Another measure of the strength of a FQHE state is the excitation gap. The
 lowest energy
 excitations of the 3/8 state are expected to lie within the spin-down CF LL,
 because of the
 reduced effective interaction.  Therefore, our effective model
 containing only spin-down composite fermions is also valid for low energy
 excitations.
 Fig.~(\ref{dispersion}) shows the low energy spectrum for 8, 10, 14, and 16
 particles
 for the effective interaction, $V^{CF}$, obtained by the Lanczos method.
 The energy gap is on the order of $\sim 0.0004 \frac{e^2}{\epsilon l_0}$.
 While the smallest gap is not identical to the gap to creating a far separated
 particle-hole pair, we expect both to be of similar magnitude. In an analogous
 study,
 Morf \cite{Morf} estimated the gap at $\nu=5/2$ to be $0.02 \frac{e^2}{\epsilon
 l_0}$.
 For a given density, the units $\frac{e^2}{\epsilon l_0}$ differ by a factor of
 $\sqrt{20/3}$ at
 5/2 and 3/8, and the theoretical estimates for the 5/2 and 3/8 gaps differ by a
 factor of
 $\sim$ 20 in constant units (e.g., mK).  Experimentally, the
 gap at 5/2 is in the range 200-300 mK,\cite{Pan,Eisenstein} which would suggest
 that the gap
 for the paired state at 3/8
 might be in the range 10-15 mK, which is quite small but above the lowest
 temperatures
 where FQHE experiments have been performed.\cite{Pan}  (Given that the 5/2 gap
 is a factor of 3-5
 smaller than the theoretical value, the number 10-15 mK ought to be taken only
 as
 a crude estimate.)

 A sufficiently large Zeeman energy will eliminate a partially polarized ground
 state at $\nu=\frac{3}{8}$.  In order to estimate the magnetic
 field range where the partially polarized state may be viable, let us consider
 the addition
 of a single composite fermion to the state in which all states of the
 0$\uparrow$ CF-LL
 are occupied.  The composite fermion can be added to either the 0$\downarrow$
 CF-LL or the
 1$\uparrow$ CF-LL.  The former is favorable so long as the Zeeman splitting
 energy,
 $\Delta_{Z}=|g|\mu_BB$, is smaller than the effective cyclotron energy of the
 composite
 fermion, $\hbar \omega_c^*$, defined
 as $\hbar \omega_c^*=\epsilon^{1\uparrow}-\epsilon^{0\downarrow}$ where
 $\epsilon^{1\uparrow}$ and
 $\epsilon^{0\downarrow}$ are the Coulomb self energies for the additional
 composite
 fermion in the 1$\uparrow$ and 0$\downarrow$
 CF-LLs.  It has been estimated from exact diagonalization as well as CF wave
 functions
 \cite{Quinn,Rezayi,Mandal} that $\hbar \omega_c^* =
 0.028 \frac{e^2}{\epsilon l_0}$ for the Coulomb interaction.
 The condition favoring the addition of the spin reversed CF, $\hbar
 \omega_c^*>\Delta_Z$, is
 satisfied for $B<20$T for parameters appropriate to GaAs.
 Many experimentally relevant effects like finite thickness, LL mixing, or
 disorder have not been considered in this estimation, and also the problem of
 the addition of
 only a single composite fermion at $\nu=1/3$ has been considered, but it
 nonetheless indicates that
 a partially polarized state at 3/8 may be possible in the accessible parameter
 regime.
 (We note that numerous partially or un-polarized FQHE states have been
 observed.\cite{CFbook})

 Before concluding, it is worth noting how fantastically complex the proposed
 3/8 paired state is when viewed in terms of electrons:  first all electrons at
 $\nu=3/8$
 capture two vortices to become $^2$CFs at $\nu^*=1+1/2$;  then those in the half
 filled spin reversed
 $^2$CF Landau level capture two additional vortices to transform into $^4$CFs
 that see no magnetic
 field;  these would normally form a $^4$CF Fermi sea,  but the Fermi sea is
 unstable to pairing due to a
 weak residual attraction between the $^4$CFs;  pairing of $^4$CFs opens up
 a gap to produce FQHE.  An observation of FQHE at this even
 denominator fraction in the lowest LL, apart from being interesting in its own
 right,
 will provide further support for pairing of composite fermions as a valid
 mechanism for FQHE.

 This work was supported in part by the National Science Foundation under grants
 no.
 DMR-9986806, DGE-9987589, and DMR-0086191 (EHR).  We are grateful to the
 Numerically Intensive
 Computing Group led by V.
 Agarwala, J. Holmes, and J. Nucciarone, at the Penn State University CAC for
 assistance and computing time with the LION-X cluster.  We thank Seung-Yeop
 Lee for many helpful discussions.

 \pagebreak

 \begin{table}
 \label{tab1}
 \caption{Overlaps of the 3/8 and 5/2 states, defined in text,
  with the paired Pfaffian wave function.}
 \begin{tabular}{|c|c|c|}
 $N$    &  $\frac{5}{2}$  & $\frac{3}{8}$ \\ \hline
 8  &  0.87 & 0.99 \\ \hline
 10 &  0.84 & 0.95 \\ \hline
 14 &  0.69 & 0.87 \\ \hline
 16 &  0.78 & 0.86 \\
 \end{tabular}
 \end{table}


\begin{references}
\bibitem{Comment2} The argument leading to the
 $2p$ enhancement of the relative angular momentum is valid only when the
 wave function requires no lowest LL projection, as is the case for
 spin reversed composite fermions in the 0$\downarrow$ $^2$CF LL.
\bibitem{comment} The Pfaffian states for $N=6$ and 12 occur at the same flux as
 the FQHE states $\nu=\frac{2}{3}$ and $\frac{3}{5}$, and will therefore
 not be considered.
\end{references}
 \end{document}